# FUV PHOTOIONIZATION OF TITAN ATMOSPHERIC AEROSOLS

Sarah Tigrine[1, 2], Nathalie Carrasco[1, 3*], Dusan K. Bozanic[2,§], Gustavo A. Garcia[2] and Laurent Nahon[2*]

[1]LATMOS/IPSL, UVSQ, Université Paris-Saclay, UPMC Univ. Paris 06, CNRS, Guyancourt, France.

[2] Synchrotron SOLEIL, l'Orme des Merisiers, St Aubin, BP48, 91192 Gif sur Yvette Cedex, France.

[3]Institut Universitaire de France, France.

§ Present address: Vinča Institute of Nuclear Sciences, University of Belgrade, P.O. Box 522, 11001 Belgrade, Serbia

* Correspondance : nathalie.carrasco@latmos.ipsl.fr, laurent.nahon@synchrotron-soleil.fr

ABSTRACT

Thanks to the Cassini Huygens mission, it is now established that the first aerosols in Titan's upper atmosphere are found from an altitude of ~1200 km. Once they are formed and through their descent towards the surface, these nanoparticles are submitted to persistent Far Ultra-Violet (FUV) radiation that can reach lower atmospheric layers. Such an interaction has an impact, especially on the chemistry and charge budget of the atmospheric compounds. Models are useful to understand this photoprocessing, but they lack important input data such as the photoemission threshold or the absolute photoabsorption/emission cross sections of the aerosols. In order to quantify the photoemission processes, analogs of Titan's aerosols have been studied with the DESIRS FUV beamline at the synchrotron SOLEIL as isolated substrate-free nanoparticles. We present here the corresponding ARPES (Angle-Resolved PhotoElectron Spectroscopy) data recorded at different FUV photon energies. The results show a very low photoionization threshold (6.0 ± 0.1 eV~ 207 nm) and very high absolute ionization cross sections (~$10^6$ Mb) indicating that FUV photoemission from aerosols is an intense source of slow electrons which has to be taken into account in photochemical models of Titan's atmosphere.

1. INTRODUCTION

Titan's atmosphere can be considered as a complex chemical reactor and as such has been the subject of both space missions and laboratory astrophysical studies. In particular, a rich and fascinating photochemistry involving Far Ultra Violet (FUV) photons (defined here as corresponding to the 200-30 nm range, i.e 6-40 eV), interacting with the main atmospheric gas phase compounds (molecular nitrogen $N_2$ and methane $CH_4$) leads to a molecular growth process with the formation of small molecular intermediates up to large solid particles (aerosols), responsible for the brownish haze surrounding Titan.

The Cassini-Huygens mission and its instruments have given some information about the aerosols' spectral signature from the far infra-red up to the FUV range (Rannou et al. 2010; Vinatier et al. 2012; Anderson & Samuelson 2011; Koskinen et al. 2011; Porco et al. 2005), but, even if progress on the FUV-induced chemical aging of the aerosols is being made (Carrasco et al. 2018), still not much is known about their FUV photodynamics once they are formed, and especially about their possible photoionization in the atmosphere.

The CAPS (Cassini Plasma Spectrometer) instrument has detected heavy negative ions ($200 \leq m/z \leq 8000 \, Da$) at an altitude of about 1000 km, with a maximum mass peak around $2000 \, Da$. Those negative ions have been identified as embryos of the aerosols that might have been formed by aggregation of PAHs (PolyAromatic Hydrocarbons) with a high electronic affinity in a ionosphere possessing a high electronic density ($n_e \approx 10^9 m^{-3}$) (Waite et al. 2007). From the moment the aerosols get charged, the main growth path will not be by coagulation anymore as the negative charges will repel each other. However, they will attract the positive ions present at the same altitude. This cation capture leads to a fast and efficient growth, starting around 1150 km and reaching a mass of 10 000 $Da$ at an altitude of 950 km. This ionic growth, led by the interaction between heavy anions and lighter cations, appears to be dominant in the upper atmosphere until sedimentation, which is responsible for the haze, begins at altitudes below 600 km (Lavvas et al. 2013). In this way, the presence of aerosols, and more importantly, their development towards nanometric sizes is mainly due to the presence of both negative and positive ions in the upper layers of the atmosphere, due to ionization of neutrals by the FUV solar light. All these processes are responsible for the presence of aerosols in the upper layers of Titan's atmosphere, starting around 1150 km.

The FUV solar light flux density at the distance of Titan (10 AU) does not exceed $10^7 ph.s^{-1}.cm^{-2}.nm^{-1}$, except for the Lyman-α band (121.6 nm) which is a hundred times more intense with a flux density of $10^9 ph.s^{-1}.cm^{-2}.nm^{-1}$ (Figure 1.A adapted from (Thuillier et al. 2004)). Since the FUV absorption cross sections and the density profiles of the main gaseous compounds ($N_2$, $CH_4$, $C_2H_2$, $C_2H_4$) are known thanks to the INMS data (Vuitton et al. 2007; Cui et al. 2009; Magee et al. 2009), we can calculate the absorption related to each of them as a function of the altitude together with the resulting total absorption.

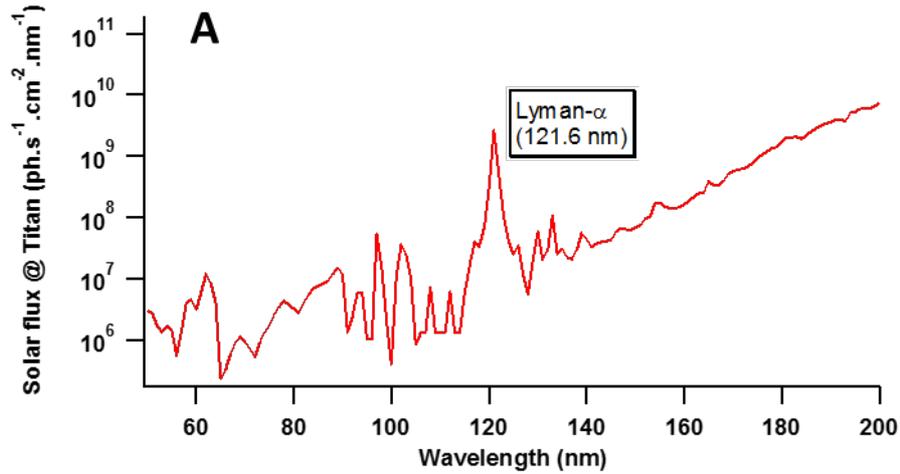

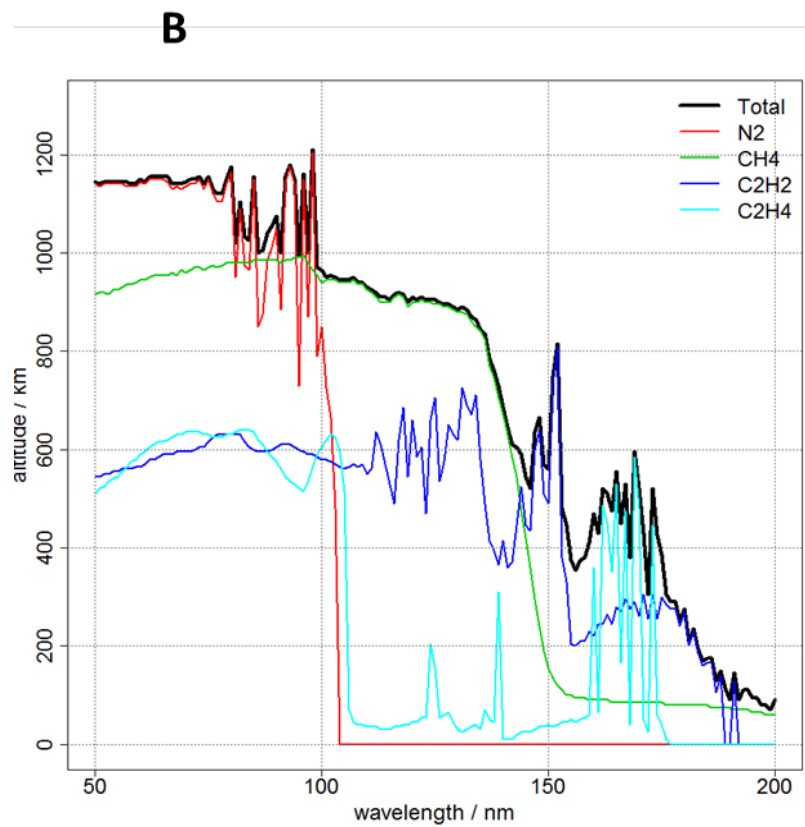

**Figure 1. (A)** Solar flux at the distance of Titan adapted from(Thuillier et al. 2004). **(B).** $\tau(\lambda) = 1$ profiles of the FUV radiations in the upper atmosphere of Titan represented by its four main chemical compounds $N_2$, $CH_4$, $C_2H_2$ and $C_2H_4$. We used their vertical density profiles together with their absorption cross sections to obtain these curves. The black curve represents the total absorption.

The penetration depth of each wavelength of the FUV solar spectrum in the atmosphere of Titan can therefore be estimated as being the altitude corresponding to an optical thickness $\tau$ such that $\tau(\lambda) = 1$ (Figure 1.B) (personal communication from P. Pernot). If we note $I_t$ the transmitted light and $I_0$ the initial intensity before passing through the atmosphere, the Beer-Lambert law states: $I_t/I_0 = e^{-\tau}$. The $\tau(\lambda) = 1$ profiles show then where the transmission is no longer significant (~37%).

As visible from Figure 1.B, established by taking into account only the absorption by the four main neutral gas phase compounds, the FUV radiation penetrates quite far down the atmosphere reaching lower altitudes. In particular, even if radiation with wavelengths below 100 nm never goes further than 900 km because of the absorption by molecular nitrogen, radiation with wavelengths between 100 and 150 nm reaches lower altitudes of 400 km. Note also that photons with wavelengths up to 200 nm penetrate down to the lower stratosphere (below 100 km).

If this FUV light is not fully absorbed by gas phase molecular species, it remains available for interaction with any other species in Titan's atmosphere, and in particular with aerosols, formed at an altitude as high as 1150 km. Therefore one should consider, in the radiative budget of Titan's atmosphere, the interaction between freshly formed aerosols that are precipitating in the atmosphere, and non-absorbed FUV radiation that penetrates down the lower atmospheric layers with wavelengths up to 200 nm.

One Titanian day lasts approximately $1.4 \times 10^6 s$ (~11 days), a time span during which the hemisphere facing the Sun is continuously irradiated by solar photons, including FUV photons. Moreover, this duration matches the residence time of the aerosols in the thermosphere, i.e. between 1000 and 600 km (Lavvas et al. 2011). Then, it is relevant to consider that the aerosols will interact with the FUV radiation ($< 200\ nm$) during their descent through this atmospheric layer.

In the present study, we are interested in the photophysical aspects of this interaction by focusing on the FUV photoionisation of the aerosols. We determine the efficiency of this process in Titan's atmosphere and its likely impact on its environment, especially regarding the charge balance. By simulating in the laboratory such a photoemission process, the minimal energy required to extract photoelectrons from the aerosols (ionization threshold), corresponding absolute photoionization cross sections and photoelectron energy distribution are measured for the first time. The later parameter appears to be crucial in order to determine if these photoelectrons are able to form a secondary energy source in the atmosphere and interact with the surrounding chemical species via electronic attachment on neutrals and electronic recombination with cations.

## 2. EXPERIMENTAL METHODS

### 2.1. Preparation of the Aerosol Samples

Our samples are analogs of the aerosols present in Titan's atmosphere, the so-called "tholins" produced within the PAMPRE dusty plasma reactor, which is described in detail in (Szopa et al. 2006). PAMPRE provides an efficient conversion of the initial $N_2$-$CH_4$ gas phase precursors, representative of the atmosphere of Titan, into a solid phase with the synthesis of tholins, which grow in suspension in the plasma. (Sciamma-O'Brien et al. 2010) showed that for a pressure inside the reactor around 1 mbar, the tholins production rate becomes optimal for an initial methane concentration of 5%, conditions that were chosen here in order to collect as much tholins as possible. PAMPRE was operated for 8 hours at a RF power of 30 W to produce 240 mg of aerosol analogs.

Moreover, several analyses have been performed in order to determine if those tholins would be representative analogs of the aerosols present in Titan's atmosphere. Concretely, different experimental results have been compared to the data gathered during the Cassini-Huygens mission. For example, (Coll et al. 2013) compared the data from the instrument Huygens-ACP (Aerosol Collector Pyrolyser) to the results of analyses performed on analogs produced by different "Titan-like" plasma experiments. More specifically, they found that "cold plasma", just like the one in PAMPRE (i.e. where neutrals remain at room temperature, with a low ionization fraction), are compatible with the composition of tholins measured by ACP, especially regarding the density of volatile species like $NH_3$ and HCN. We can therefore consider the nanoparticles (NPs) produced by the PAMPRE platform as relevant analogs of the aerosols present in Titan's upper atmosphere.

*2.2 Aerosol Production and Characterization*

Free tholins nanoparticles were produced by nebulizing an 1 $g.l^{-1}$ aqueous solution of the above-described tholins in an atomizer (TSI, model 3076), using 2 bars of $N_2$ as the atomizing gas. These NPs were subsequently dried in a 2 m–long (TSI, model 3062) diffusion dryer. In order to fully characterize our aerosol analogs, we measured their size distribution with a SMPS (Scanning Mobility Particle Sizer) device available on the DESIRS beamline (model TSI 3080L/3772) as discussed in (Gaie-Levrel et al. 2011).

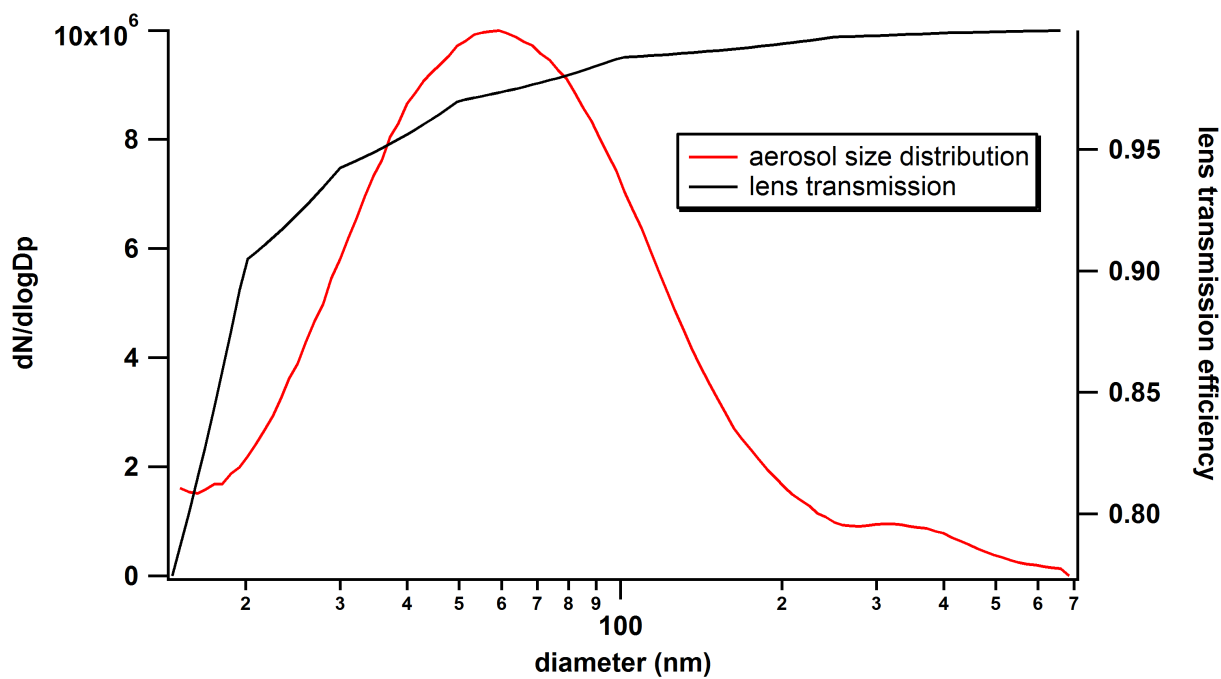

**Figure 2. Size distribution of the tholins obtained with a SMPS showing a bimodal structure with two main peaks centered at 60 and 325 nm (red curve). The black curve represents the transmission of the aerodynamic lens for each nanoparticle diameter.**

Figure 2, left and right axis, shows respectively the experimentally measured SMPS count size distribution of the airborne tholins and the theoretical (Wang et al. 2005) transmission efficiency of the aerodynamic lens system (see Sec. 2.4). A bimodal size distribution, with two modal diameters at 60 and 325 nm, was measured with number concentrations of $1.5\times10^5$ and $2.1\times10^4$ $particles.cm^{-3}$ at their maxima, respectively. By taking into account the transmission efficiency, 97% and 99% of these first and second modes respectively are theoretically transmitted by the lens system with an overall transmission of 96 % of the total distribution.

A second method based upon a scanning electron microscope (SEM, ZEISS Supra55VP) was also used to characterize the produced tholins particles. Two types of samplings were performed: (1) - deposition/evaporation of a colloidal solution droplet on a SEM grid, (2) - electrostatic precipitation of airborne tholins particles using a nanometer aerosol sampler (NAS, model 3089, TSI) after the nebulization process (Figure 3 a,b).

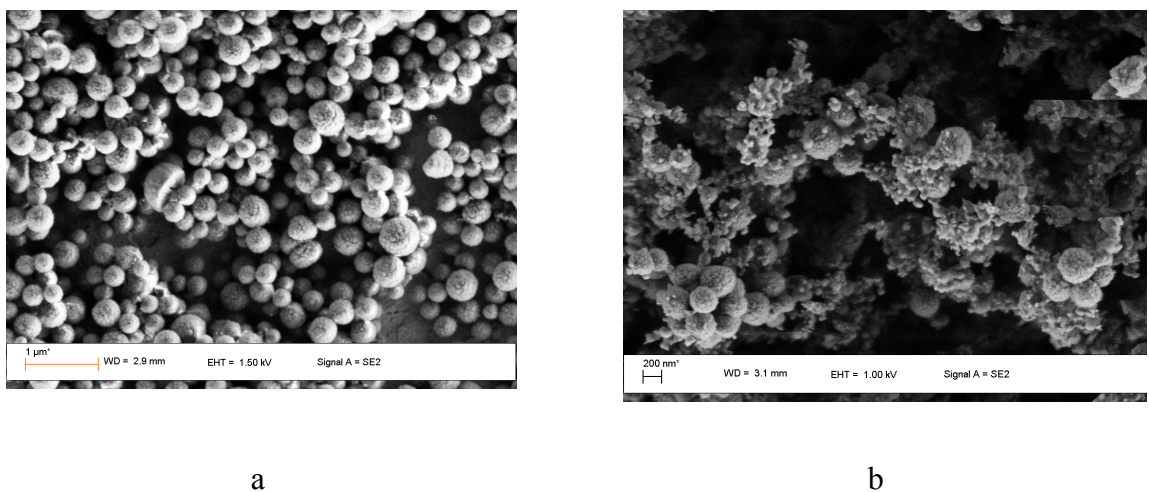

a　　　　　　　　　　　　　　　　　　b

Figure 3 : Scanning Electron Microscope (SEM) images of tholins particles, (a) in the colloidal suspension (1µm scale), (b) after the aerosol production (200 nm scale).

Figure 3(a) presents a SEM picture obtained for the first type of sampling. A count size distribution with a 350 nm modal diameter was measured and corroborates the second mode seen on the SMPS for the same PAMPRE experimental conditions. An earlier study of (Hadamcik et al. 2009) showed that, in this reactor, the average grain size increases with the amount of $CH_4$ and, more generally, depends on the plasma parameters such as the gas flow, the discharge power and the duration of the experiment. As an example, a 300 nm mode diameter was obtained for a gaseous mixture containing 5% of methane in a 30 W RF discharge, which is consistent with our measurements. On the other hand, Figure 3(b) presents a SEM picture obtained for the second type of sampling and confirmed the SMPS bimodal distribution with modal diameters of 50 and 345 nm. The first mode at 50 nm was produced during the nebulization process and could correspond to the SMPS peak at 60 nm. Because the additional family of smaller particles seen in Figure 2 and Figure **3**(b)—with diameters in the 50-60 nm range—appear only when going through the nebulization process, we believe that they are most likely a consequence of the violent nature of the atomization. In all cases in the following we will consider that the size distribution of the NPs entering the experimental vacuum chamber is the one given by Figure 2, with 94% of the particles falling into the first mode with a mean size of 60 nm and the 6% remaining onto the second mode at 325 nm.

Finally, note that we used water as a solvent for the tholins prior to their nebulization. Besides being the one with the lowest evaporation rate, water has the highest ionization threshold (12.6 eV against, for example, 10.8 eV for methanol or 12.2 eV for acetonitrile). Therefore, possible residual gas phase water molecules surrounding the tholins are fully transparent up to 12.6 eV.

*2.3. The FUV Ionizing Radiation: the DESIRS Beamline at Synchrotron SOLEIL*

Intense tunable synchrotron FUV light was produced by the undulator-based beamline DESIRS at Synchrotron SOLEIL (Gif sur Yvette, France) delivering photons of high brightness in the 4 to 40 eV range with variable polarization (Nahon et al. 2012). We used a moderate resolution grating (200 groove/mm) providing flux in the $10^{12} - 10^{13}\, ph.s^{-1}$ range for a bandwidth $\Delta\lambda/\lambda$ ~0.1%. Note that, thanks to a rare gas filter system, acting as a low-energy pass filter, the radiation provided by DESIRS is free of any undulator-harmonics.

*2.4. The Photoionization Chamber and the Photoelectron Spectrometer*

Experiments were performed at the SAPHIRS molecular beam endstation (Tang et al. 2015). The exact same method as described in Sec 2.2 was used to produce free tholins nanoparticles by nebulization followed by a drying stage before entering, through a 190 μm limiting aperture, an aerodynamic lens (Gaie-Levrel et al. 2011) that focuses them into a ~ 400 μm diameter aerosol beam with a very high transmission (above 90 % for nanoparticles ranging from 20 nm to 1500 nm). The nanoparticles then cross two ϕ=2mm skimmers before reaching the center of the ionization region. Note that the SAPHIRS double-differential vacuum system efficiently removes the gas phase component of the beam (mostly $N_2$ and $H_2O$ in this case since tholins themselves have an extremely low vapor pressure), so that most of the electron signal comes from the nanoparticle phase.

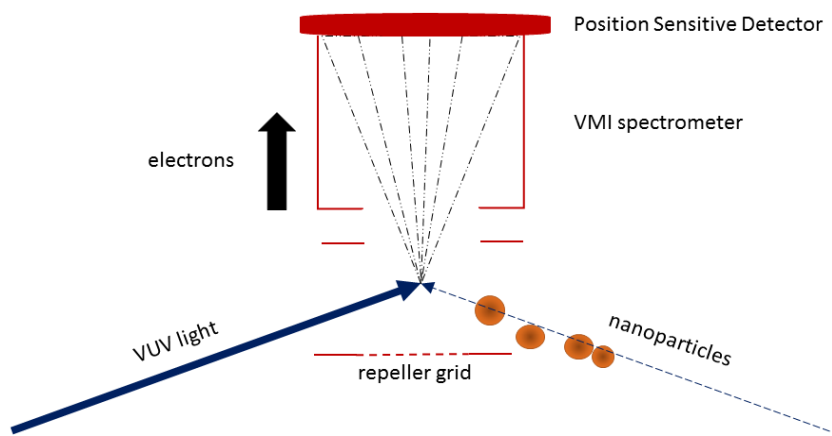

Once in the interaction zone, the tholins are photoionized by the FUV synchrotron radiation (Figure 4) releasing a photoelectron with a kinetic energy $E_k$ such that $hv = E_i + E_k$ where $hv$ is the incident energy of the FUV photons and $E_i$ the ionization energy of a given electronic orbital. Photoelectron Spectroscopy (PES) consists in measuring the kinetic energy distribution of the photoelectrons at a given photon energy, which in turn yields the spectroscopy of the cation, i.e. its ionization energies. To do so, we used the double imaging electron/ion coincidence spectrometer (i$^2$PEPICO) DELICIOUS 3 (Garcia et al. 2013) which possesses on the electron side a Velocity Map Imaging spectrometer (VMI) (Eppink & Parker 1997) in which the 3D expending sphere of photoelectrons is imaged onto a 2D position sensitive detector (PSD) providing multiplex radial and angular information. The electrostatic lens formed in the interaction region is such that the impact location onto the PSD only depends on the initial momentum of the electrons and not on their precise location of creation. The radius of the circular pattern obtained on the PSD depends on $E_k$ and the electrostatic potentials, which can of course be tuned so that the 4π distribution of electron is collected. Provided that there is an axis of symmetry in the 3D photoelectron angular distribution contained in the detector plane, the 2D image can be Abel-inversed to retrieve a central cut of the 3D distribution whose angular integration directly leads to the PES.

**Figure 4. Illustration of the nanoparticle ionization setup used on the SAPHIRS platform on the DESIRS beamline. The aerodynamic lens injects free NPs in the ionization chamber under vacuum, where they cross at right angle the FUV synchrotron radiation. The produced electrons are guided and momentum-focused onto a position sensitive detectors (PSD) via a VMI spectrometer.**

In practice, the VMI concept has already been used to study the angle-resolved PES of salt NPs (Wilson et al. 2007; Goldmann et al. 2015) but only raw images where presented without any image inversion yielding the exact PES. This is due to the fact that in these previous works there was no revolution symmetry axis in the 3D electron distribution. Indeed, the electric field associated with the linear polarization of the FUV light which was used in these studies was perpendicular to the photon propagation axis along which a forward/backward asymmetry is observed in NP photoemission (as seen on Figure 5). This well-documented photon-dependent asymmetry is due to an interplay between the NP size dependent finite photon penetration depth and electron escape depth. In our case, thanks to the circularly polarized light (CPL) available on DESIRS, the quantification axis is the propagation axis and the revolution symmetry requirement is fulfilled, so that electron images can be Abel inverted. To do so we used the Basex method (Dribinski et al. 2002) providing both radial and angular information, i.e. Angle-Resolved Photoelectron Spectroscopy (AR-PES). We focus, in the present study, onto the analysis of the radial distribution, by angular integration of the inverted images, providing the PES, i.e the photoelectron kinetic energy distribution, with an overall relative energy resolution of about 4 %.

At each photon energy, the PES experiment on NPs is followed by a similar "background" measurement with the water solvent alone to quantify and then subtract its residual contribution, in order to provide the pure signal from the tholins only.

3. RESULTS AND DISCUSSION

We have used different FUV photon energies from 9 to 11 eV to photoionize the PAMPRE tholins (whose ionization energy was a priori unknown), including of course the $121.6\ nm$ ($10.2\ eV$) Lyman-α emission line, which is the most intense one in the FUV solar spectrum (a hundred times higher than the average, see Figure 1.A).

The upper limit in energy is set by the activation threshold ($100\ nm = 12.3\ eV$) of molecular nitrogen: in the atmosphere of Titan, we assume that below 100 nm, the light will be mainly absorbed by the major gas phase compounds $N_2$ and $CH_4$ and will marginally interact with the aerosols. As mentioned earlier after the 11 day solar irradiation time the aerosols have descended to an altitude of 600 km, which corresponds to a lower energy limit of 150 nm (Figure 1.B) Therefore, the energy range presented in this work, between 9 and 11 eV, represents the photoprocessing of aerosols outside the absorption spectrum of $N_2$, from their formation at high altitudes out up to the end of their diurnal continuous irradiation in the thermosphere.

In Figure 5 are shown the background-corrected non-inverted images obtained at various photon energies, from which are extracted, after inversion, the results presented and discussed hereafter. It is obvious to the naked eye that the forward/backward asymmetry increases with photon energy because the particles become more opaque at wavelengths closer to their diameter (Wilson et al. 2007). There is a relation between angular distributions and physical and optical properties of nanoparticles (Wilson et al. 2007; Goldmann et al. 2015; Signorell et al. 2016; Jacobs et al. 2017). The analysis of this angular information in the case of tholins will be the subject of a forthcoming paper.

**Figure 5.** Corrected velocity map images obtained at photon energies: 9 eV (top, left), 9.5 eV (top, right), 10.2 eV (bottom, left) and 11 eV (bottom, right). Note the forward/backward asymmetry observed in the NP photoemission along the light propagation axis (red arrow).

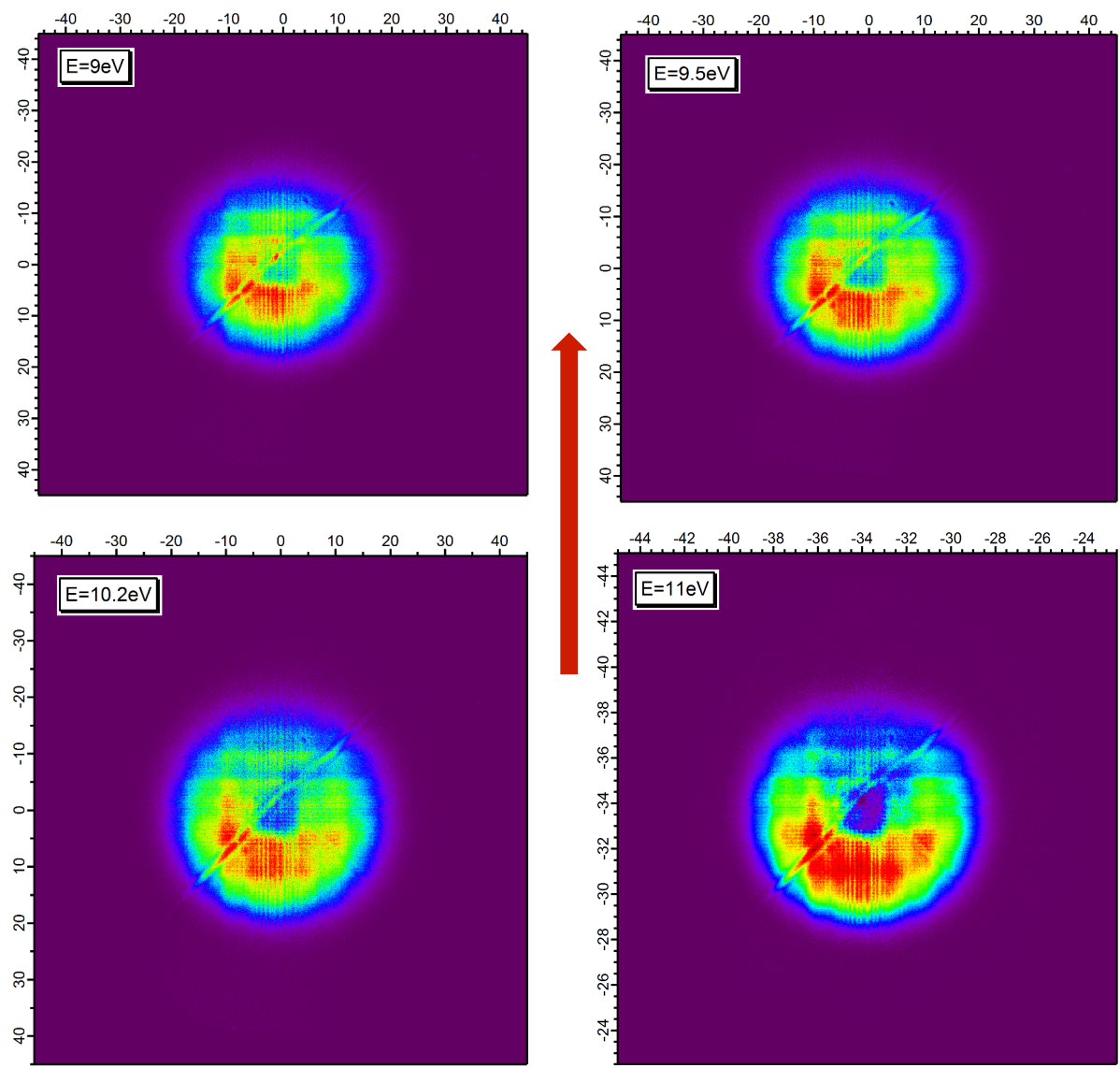

*3.1 The Ionization Threshold*

Figure 6 shows the photoelectron spectra (PES) obtained by Abel inversion of the corrected images (Figure 5) and angular integration at different photon energies. We notice that the ionization onset, as estimated by the departure from the baseline signal, appears at a photon energy of $6.0 \pm 0.1\,eV$ ($\sim 207\,nm$) regardless of the incident energy. This corresponds then to the ionization threshold of the tholins composed of 5% methane, determined here experimentally and directly for the first time for this kind of materials. This means that any photon whose energy is above 6 eV may interact with the tholins leading to their ionization. This value has to be compared to the ionization threshold of the main surrounding neutral gas phase compounds in Titan atmosphere, which are well above: $N_2$ at 15.6 eV, $CH_4$ at 12.6 eV, $C_2H_2$ at 11.4 eV, and $C_2H_4$ at 10.5 eV.

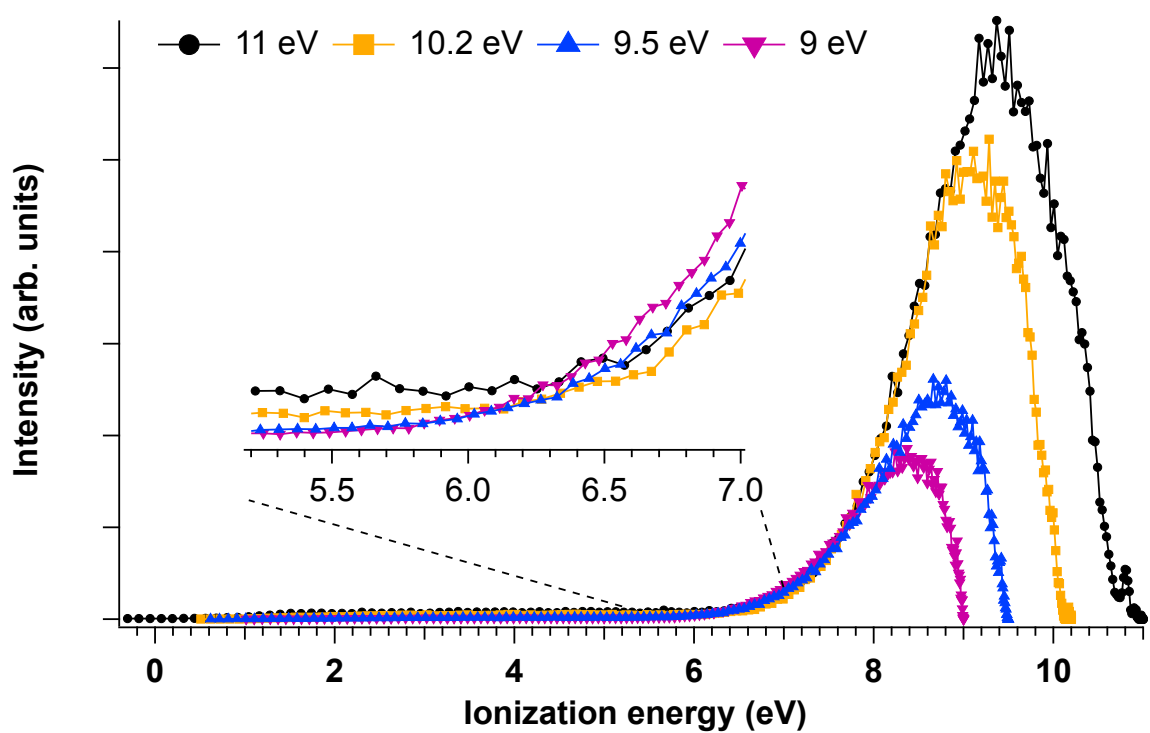

**Figure 6.** Photoelectron spectra of tholins aerosols recorded at different photon energies (11, 10.2, 9.5 and 9 eV). The area of the curves have been normalized to the total cross-sections derived in Sec. 3.3. The inset shows a zoom of the ionization onset.

This is consistent with the predictions of Borucki & Whitten, who state in their study, (Borucki & Whitten 2008) that the FUV ionization of the aerosols has to start at an energy lower than 7 eV in order to be effective, while higher energies would be absorbed by the gas phase. (Mishra et al. 2014) went further in their analysis by making the threshold vary in order to fit the data of the PWA (Permittivity, Waves and Altimetry) instrument onboard Huygens between 160 and 80 km in the atmosphere suggesting that, as the nature of the aerosols changes during their descent through the atmosphere (size, chemical composition etc...), their ionization threshold might change as well and may vary as a function of the altitude. They managed to satisfactorily reproduce the data obtained by the PWA permittivity instrument by varying the ionization threshold from 6 to 7 eV. Our findings are consistent with their lower limit energy range. Our experimentally measured value is however significantly higher than the ~ 5.1 eV calculated for 3 nm diameter carbonaceous aerosols in (Lavvas et al. 2013). This discrepancy might be due to the large amount of nitrogen in our analogs (Sciamma-O'Brien et al. 2010) as compared to PAH nanoparticles.

Note that this photoionization threshold measured here for neutral tholins is most probably very similar for charged tholins. Indeed for large molecular structures, above a size of few nanometers, the charge density is so low that the ionization threshold does not depend on the initial charge (Giuliani et al. 2012).

*3.2 Kinetic Energy Distribution of the Photoelectrons*

In order to analyze phenomenologically the photoelectron kinetic energy distributions shown in Figure 7, we have extracted the position of the most probable kinetic energy and the Full Width at Half-Maximum (FWHM) of the PES and these are shown as an insert in the same figure. As the photon energy increases, the most probable kinetic energy shifts towards higher values: from 0.6 $eV$ at $hv = 9\ eV$ to 1.5 $eV$ for $hv = 11\ eV$. In parallel, the FWHM of the distribution increases too, from a FWHM of ~1.2 eV at $hv = 9\ eV$, to ~1.9 eV at $hv = 11\ eV$. Both trends are expected since an increase in the photon energy leads to more energetic electrons, on one hand, and to the widening of the energy distribution as more orbitals are being ionized, on the other hand.

Finally, at 11 eV, the kinetic energy distribution appears to have a dual component with in addition to the fast peak centered around 2 eV, a second weaker slow component around 0.8 eV. The latter contribution is due to electrons that lose energy by inelastic scattering prior to photoemission (Shu et al. 2006; Goldmann et al. 2015), whereas the band observed at higher kinetic energies is the result of the direct photoemission without scattering. This 'secondary' photoemission is also present in the PES obtained for lower photon energies, but the direct photoelectrons are not fast enough to induce a clear separation between both contributions below $hv = 11\ eV$. Note that the secondary photoemission in the FUV domain is a consequence of finite, yet small, size of the aerosol particles. More specifically, in this configuration the electron-electron scattering length and the absorption length values are comparable with the particle size, leading to commensurate probabilities for both secondary and direct photoemission.

For a given photon energy, a low value of the ionization threshold allows the produced photoelectrons to acquire more kinetic energy as shown above. This KE makes them mobile in the atmosphere, especially because they are lighter as compared to the surrounding ions. However, their KE is low enough so that they can recombine efficiently with aerosols even if the maximum number of electrons that a nanoparticle can receive depends, generally, on its size, and therefore on the altitude. (Lavvas et al. 2013) focused, for example, on the ionization peak at the altitude of 1100 km and found that the aerosols found there are mainly negatively charged, with the ability to capture only one free electrons. Moreover, they showed that below the altitude of 1000 km, the electron density decreases while, in parallel, the amount of negatively charged aerosols increases. This effect goes on with decreasing altitude as the aerosols become larger, leading to an increased cross section for electron capture.

In addition, those rather "slow" photoelectrons, with a kinetic energy below 10 eV, are ideal candidates for dissociative recombination reactions with surrounding gas phase cations or dissociative electronic attachment reactions on neutrals. Those reactions have high cross sections for electrons with energies $\leq 10\ eV$ (Dobrijevic et al. 2016) and need to be considered in order to understand the atmospheric densities of neutrals (Plessis et al. 2012) and negative ions.

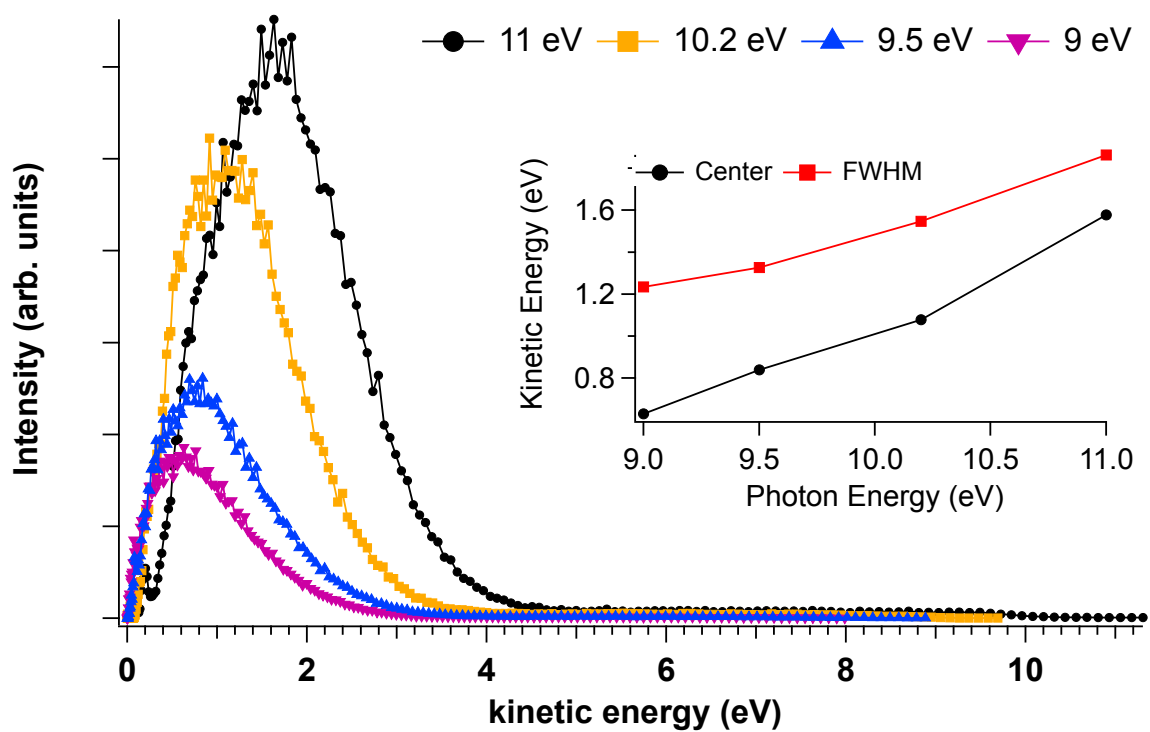

**Figure 7. Kinetic energy distributions of the electrons coming from photoionization of tholins at different photon energies. The area of the curves have been normalized to the total cross-sections derived in Sec. 3.3. The inset shows the evolution of the position of the most probable kinetic energy (center) and FWHM as a function of the photon energy.**

*3.3 The Absolute Ionization Cross Sections*

Another crucial parameter to determine the efficiency of the tholins FUV photoionization is the ionization cross sections and their photon energy dependence. By definition, the ionization cross section σ is given by:

$$\sigma(\lambda) = \frac{photoelectron\ rate\ (\lambda)}{photon\ flux\ (\lambda) \times nanoparticle\ density \times interaction\ length}$$

where the photoelectron rate (in $s^{-1}$) is the total amount of photoelectrons collected divided by the total duration of a given measurement corrected by the 60 % detector efficiency (corresponding to the open area of the microchannel plate detectors); the interaction length corresponds to the transversal section of the (cylindrical) nanoparticle beam which has been measured by laser Mie scattering (Gaie-Levrel et al. 2011) to be 420 μm.

The particle concentration is estimated from the convolution of the size distribution and the size-dependent transmission of the aerodynamic lens to $3.5 \times 10^6\ particles.cm^{-3}$. Considering a matter flow via the aerodynamic lens pinholes of $3.17\ cm^3.s^{-1}$ one infers a flow of $1.37 \times 10^7\ particles.s^{-1}$ entering the expansion chamber. The aerosol beam volume of $25\ cm^3$ is estimated from a cylinder of $420\ \mu m$ diameter and length equal to the beam's speed, which simulated value is $180\ m.s^{-1}$, leading to a particle density of $5.5 \times 10^5\ particles.cm^{-3}$ at the ionization region, if one assumes a 100 % transmission via the 2-mm diameter skimmers.

For our measurement conditions, the photon flux of the DESIRS beamline has been determined by using a calibrated Si photodiode (IRD, AXUV 100), and ranged from $1.1 \times 10^{12}$ to $7.6 \times 10^{12}\ photon.s^{-1}$.

The calculated cross sections, taking into account all the above-described parameters are given in Mbarn ($10^{-22}\ m^2$) in Figure 8. The values start at $2.4 \times 10^5\ Mb$ at 9 eV and increase to reach $1.2 \times 10^6$ *Mb* at 11 eV. For comparison, the photoionization cross sections of atoms and small molecules, in the few eV range above their ionization threshold, are in general, between a few Mb and a few tens of Mb. For instance, these values are between 20 and 25 Mb for molecular nitrogen $N_2$ (Gallagher et al. 1988), or 10 and 25 Mb for methane (Wang et al. 2008). Note that all these integrated cross sections have been used to normalize the photoelectron kinetic energy distributions presented in Figure 7. Besides, experimental uncertainties relating to tholins partial optical opacity, production and transmission, electron detection efficiency and photon flux measurements would have a proportional impact on these results. They are hard to quantify, but they would be most likely overestimated, meaning that the absolute cross-sections reported here should be taken as lower limits. In all cases our measured values correspond to a bimodal size distribution of our aerosols (peaked at 60 and 325 nm in diameter) as shown in Figure 2. Cross sections derived from a simple Mie model strongly depend on the size of the nanoparticles. However the size of our analogs are consistent with the size predicted at a 600 km altitude in Titan's atmosphere, where 6 eV radiation is still strongly present (see Figure 1).

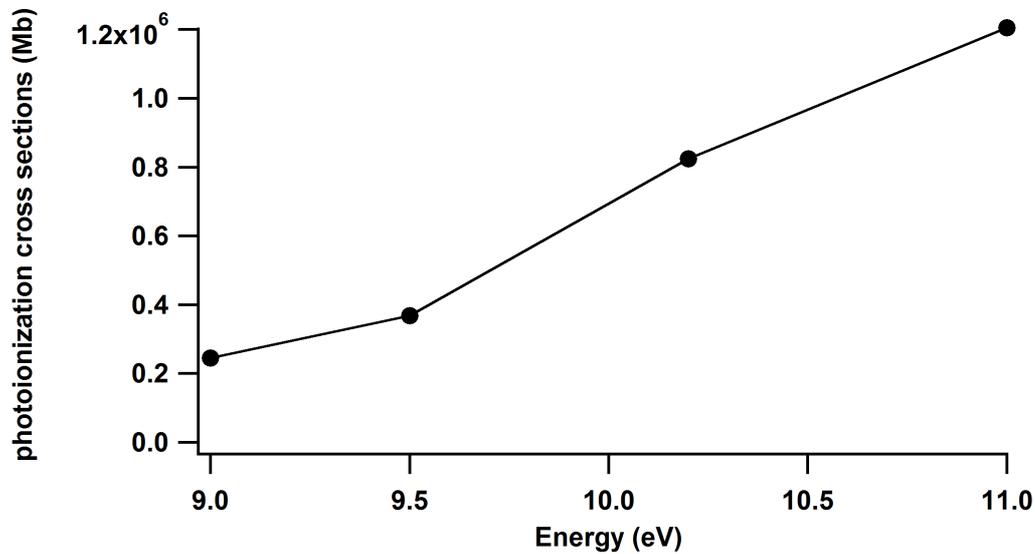

**Figure 8. Experimentally-derived absolute photoionization cross sections of the tholins in the near FUV range.**

The tholins photoionization cross section appears therefore five orders of magnitude higher than the ones of standard molecules. This result, combined with a low ionization threshold at $6.0 \pm 0.1\ eV$, make the aerosols FUV photoionization a non-negligible phenomenon in the atmosphere of Titan (as compared to X-ray photoionization for example) and should definitely be taken into account in the overall Titan atmosphere radiative transfer budget.

4. CONCLUSION

The laboratory experiments of FUV photoionization on Titan's aerosol analogs reported in this paper constitute the first experimental values for the ionization threshold, ionization cross sections and kinetic energy distributions of the corresponding photoelectrons. Due to their low ionization threshold of $6.0 \pm 0.1\ eV$, associated to very large cross sections, five orders of magnitude higher than typical gas phase molecules, FUV photoionization of aerosols is an important phenomenon and represents a non-negligible source of electrons in the atmosphere that needs to be taken into account in electrical conductivity models. In terms of chemical reactivity, the produced photoelectrons possess a kinetic energy lower than 10 eV and will either recombine with the aerosols from which they originate, heat their environment (by electronic diffusion) or initiate dissociative recombination reactions that become very efficient for low-energy electrons.

This experiment opens opportunities to improve our understanding of planetary aerosols in combination with models, by expanding the range of experimental parameters representative of the Titan atmosphere. One of these parameters is the composition of the tholins, which have been produced here from an initial mixture of $95\%\ N_2 + 5\%\ CH_4$. This ratio could be varied since the tholins composition may impact their photoemission properties. A "coating" made of ethane or methane might as well have some effect, such as increasing the ionization threshold for example.

Another crucial parameter is the size of the aerosols, which is, as the FUV spectrum, altitude-dependent. One of the caveats of our experiment is that so far we could not pre-select their size before transmission and ionization due to the low signal, so that the full size distribution of Fig. 2 contributes to the current results. Models appear to be quite sensitive to the size of the aerosols, as it might change their electronic behavior since the larger the nanoparticles the more surface they offer for interaction with the FUV in order to emit electrons (or capture them). The aerosol radius of Titan's aerosol is about 10 nm at 600 km (Lavvas et al. 2011), which is significantly smaller than the radius of the aerosol samples synthesized in our study. In order to infer the effective electron production from Titan's aerosols VUV photoionization in Titan's ionosphere, a further determination of the aerosols total VUV absorption cross section would be necessary as a normalization scaling factor for the electron production rate obtained in the present study. For future experiments, it would also be interesting to size-select the nanoparticles and play with their composition/coating in order to test the experimental sensitivity of these parameters to be compared to photochemical models and the permittivity measurements in Titan's atmosphere. Recent improvements in aerosol transmission at SAPHIRS will allow exploration of the size dependence.

## ACKNOWLEDGMENTS

We would like to thank P. Pernot, D. Dubois and F. Gaie-Levrel for helpful discussions. S. Tigrine is grateful to the Paris Saclay IDEX for providing a thesis grant under the IDI program. N. Carrasco thanks the European Research Council for funding via the ERC PrimChem project (grant agreement No. 636829). We are indebted to JF Gil for his technical help around the SAPHIRS platform and to the general technical staff of Synchrotron SOLEIL for running smoothly the facility under projects 99150054 and 99150112.

2015, J Chem Phys, 142, 0

Hadamcik, E., Renard, J. B., Alcouffe, G., et al. 2009, Planet Space Sci, 57 (Elsevier), 1631, http://dx.doi.org/10.1016/j.pss.2009.06.013

Jacobs, M. I., Kostko, O., Ahmed, M., & Wilson, K. R. 2017, Phys Chem Chem Phys, 19 (Royal Society of Chemistry), 13372

Koskinen, T. T., Yelle, R. V, Snowden, D. S., et al. 2011, Icarus, 216 (Elsevier), 507

Lavvas, P., Galand, M., Yelle, R. V., et al. 2011, Icarus, 213 (Elsevier Inc.), 233, http://dx.doi.org/10.1016/j.icarus.2011.03.001

Lavvas, P., Yelle, R. V, Koskinen, T., et al. 2013, Proc Natl Acad Sci, 110, 2729, http://www.pnas.org/content/110/8/2729.short%5Cnhttp://www.pnas.org/cgi/doi/10.1073/pnas.1217059110

Magee, B. A., Waite, J. H., Mandt, K. E., et al. 2009, Planet Space Sci, 57 (Elsevier), 1895, http://dx.doi.org/10.1016/j.pss.2009.06.016

Mishra, A., Michael, M., Tripathi, S. N., & Béghin, C. 2014, Icarus, 238, 230

Nahon, L., De Oliveira, N., Garcia, G. A., et al. 2012, J Synchrotron Radiat, 19, 508

Plessis, S., Carrasco, N., Dobrijevic, M., & Pernot, P. 2012, Icarus, 219 (Elsevier Inc.), 254, http://dx.doi.org/10.1016/j.icarus.2012.02.032

Porco, C. C., Baker, E., Barbara, J., et al. 2005, Nature, 434 (Nature Publishing Group), 159

Rannou, P., Cours, T., Le Mouélic, S., et al. 2010, Icarus, 208 (Elsevier), 850

Sciamma-O'Brien, E., Carrasco, N., Szopa, C., Buch, A., & Cernogora, G. 2010, Icarus, 209 (Elsevier Inc.), 704, http://dx.doi.org/10.1016/j.icarus.2010.04.009